\documentstyle[12pt,graphicx]{article}
\begin{document}

\title{\Large Excited states of beryllium atom from
explicitly correlated wave functions.}
\author{F.J. G\'alvez
and E. Buend\'{\i}a\\ Departamento 
de F\'{\i}sica Moderna,
Facultad de Ciencias, \\ Universidad de 
Granada, E-18071
Granada, Spain\\
A. Sarsa\\
International School for Advanced
Studies, SISSA \\
Via Beirut 2/4,  
I-34014 Trieste,
Italy
}
\date{ }
\maketitle

\begin{abstract}

A study of the first excited states of beryllium atom starting from
explicitly correlated wave functions is carried out.  Several
properties are obtained and discussed focusing on the analysis of the
Hund's rules in terms of the single--particle and electron pair
intracule and extracule densities. A systematic study of the
differences on the electronic distributions of the singlet and triplet
states is carried out. The trial wave function used to describe the different
bound states  consists of a generalized Jastrow-type correlation factor 
times a configuration interaction  model wave function. 
This model wave function has been fixed by using a generalization of the
optimized effective potential method to deal with multiconfiguration wave
functions. The
optimization of the wave function and the calculation of the different
quantities is carried out by means of the Variational Monte Carlo
method.

\end{abstract}

\newpage

{\bf I. INTRODUCTION} \\

Most of the studies on the excited states of the beryllium atom have
been focused on the attainment of the binding energy.  They have been
mainly carried out by using extensive Configuration Interaction type
wave functions obtaining very accurate results \cite{graham}.
Recently very precise values for the energy and related properties
such as the electron affinity or mass shifts have been calculated by
making use of the Quantum Monte Carlo method \cite{mella} and by using
large expansions of exponentially correlated Gaussian functions 
\cite{komasa2}.  However the knowledge of other interesting properties
of these states is much more scarce. For example, one and two body
electron densities, which provide valuable information on the
structure and dynamics of the system are not fully characterized. This
is due to the technical difficulties appearing in their calculation.
These densities play an important role in the understanding and in the
interpretation of some interesting features of the electronic
structure of atoms such as the Coulomb hole and the Hund's rule
\cite{could61}--\cite{abgp96}. This rule states that if two
states arise from the same configuration, the state having the higher
spin will have the lower energy.  For those states with the same spin
value, the most bound one is that with the higher value of the orbital
angular momentum.  Contrary to qualitative arguments, the
electron--electron repulsion is higher for the triplet state, and is
the deeper electron--nuclear attraction energy value of the triplet
state the responsible for the lower triplet energy. This has been
shown for some excited states of Helium--like systems by using
Hylleraas type wave functions \cite{regier84}, for the $^1P$--$^3P$
terms arising from the $2s$--$2p$ configuration in beryllium, by using
both Hartree-Fock and Configuration Interaction wave functions
\cite{hartree}--\cite{koga1}, and also for some other atomic and
molecular systems within the Hartree--Fock framework
\cite{koga2,boyd87}.  \\

The aim of this work is to extend previous studies to the first
excited states of the beryllium atom with symmetry $^1S$, $^3S$,
$^1P$, $^3P$, $^1D$ and $^3D$ by using explicitly correlated wave
functions.  Starting from those wave functions the energy and some
different one-- and two--body position and momentum properties as well
as the single-particle density and both the interelectronic
(intracule) and the center of mass (extracule) two--body densities in
position space have been obtained.  Explicitly correlated wave
functions fulfill some analytically known properties of the densities,
as for example the correct short range behavior and cusp conditions
which become very important in describing correctly the Coulomb hole.
These distribution functions provide a picture of the spatial
arrangement of the electrons and will be analyzed in order to get
insight on the correlation effects and on the Hund's rule.\\

The single--particle density is defined as

\begin{equation}
\rho(\vec{r}) =
\int \sum_{i=1}^{N} \delta [\vec{r} - \vec{r}_{i} ]
\frac{|\Psi(\vec{r}_{1},\vec{r}_{2},...,\vec{r}_{N})|^2}{\langle
\Psi|\Psi\rangle} d \vec{r}_{1} d \vec{r}_{2} ...d \vec{r}_{N},
\end{equation}

the interelectronic or intracule density is given by

\begin{equation}
I(\vec{r}_{12}) = 
\int
\sum_{i>j=1}^{N} \delta [\vec{r}_{12} - (\vec{r}_{i} -
\vec{r}_{j}) ]
\frac{|\Psi(\vec{r}_{1},\vec{r}_{2},...,\vec{r}_{N})|^2}{\langle
\Psi|\Psi\rangle}
d\vec{r}_{1}  d \vec{r}_{2}  ...d \vec{r}_{N}
\end{equation}
 
and, finally, the center of mass or extracule density can be
calculated as

\begin{equation}
E(\vec{R}) = \int 
\sum_{i>j=1}^{N}
\delta [\vec{R} - (\vec{r}_i + \vec{r}_j)/2]
\frac{|\Psi(\vec{r}_{1},\vec{r}_{2},...,\vec{r}_{N})|^2}{\langle
\Psi|\Psi\rangle} d\vec{r}_{1}  d \vec{r}_{2}  ...d \vec{r}_{N}  
\label{extrp}
\end{equation}

The corresponding spherically averaged densities are labelled as
$\rho(r)$, $h(r_{12})$ and $d(R)$, respectively, and represent the
probability density functions for the electron--nucleus distance, $r$,
the interelectronic distance, $r_{12}$, and the center of mass
distance, $R$.  The single--particle density is normalized to the
number of electrons, $N$, and both the intracule and the extracule
densities to the number of electron pairs, $N(N-1)/2$. The value of
the intracule and extracule densities at the origin give us the
probability of two electrons to be at the same position and at opposite
positions with respect to the nucleus, respectively.\\

The wave functions used in this work include explicitly the electronic
correlations by means of a Jastrow-type factor. Nondynamic effects are
also taken into account by using a multi configuration model wave
function where the most important substitutions are included. This
parameterization has shown to provide accurate results for the ground
state of the isoelectronic series of the beryllium atom
\cite{gbs99}. The calculation of the different expectation values has
been carried out by means of the Variational Monte Carlo method
\cite{hamm}. \\

The structure of this work is as follows. In Sec. II the trial wave
function is shown in detail. The results are presented and discussed
in Sec. III.  The conclusions of this work can be
found in Sec. IV. Atomic units are used throughout unless otherwise
indicated.\\

{\bf II. WAVE FUNCTION}\\

The correlated trial wave function, $\Psi$, used in this work is the
product of a symmetric correlation factor, $F$, which includes the
dynamic correlation among the electrons, times a model wave function,
$\Phi$, that provides the correct properties of the exact wave
function such as the spin and the angular momentum of the atom, and is
antisymmetric in the electronic coordinates.

\begin{equation}
\Psi=F \Phi
\end{equation}

For the correlation factor we use the form of Boys and Handy
\cite{BoHa1} with the prescription proposed by Schmidt and Moskowitz
\cite{ScMo1}. In particular we have worked with $17$ variational
non--linear parameters in the correlation factor which include
electron--nucleus, electron--electron and electron--electron--nucleus
correlations.\\

In order to get the different bound states of a given $L$, $S$
symmetry we make use of the well known Hylleraas-Undheim
theorem.  The hamiltonian is diagonalized in a set of $m$ linear
independent trial functions with the proper symmetry and then the
eigenvalues constitute upper bounds to the first $m$ bound states. The
trial wave functions are linear combinations of Slater determinants
built starting from the single particle configurations $1s^2nln'l'$,
with $n,n'=2,3$ and $l,l'=s,p,d$, coupled to the total orbital angular
momentum and spin of the term under study times the Jastrow
correlation factor aforementioned.  For the $^1D$ state we have also
considered the configuration $1s^22s4d$.\\

The model wave function has been fixed within the optimized effective
potential framework. This is a mean field approximation to the many
electron problem based on finding the best local one-body potential
minimizing the total energy, which is expressed as an orbital
functional as in the 
Hartree-Fock approach. This method was first proposed by Slater
\cite{slater51} as a simplification of the Hartree-Fock equations and
further developed and applied to atomic problems by Talman and
coworkers \cite{talman,talman1}.  The trial wave function is a Slater
determinant with single--particle wave functions calculated from a
certain effective potential that is assumed to be central. The minimum
condition on the effective potential leads to a set of non-linear
coupled integro-differential equations involving the potential and the
orbitals.  They can be solved by means of a self-consistent procedure
just like in the Hartree-Fock method by using either numerical-grid
techniques \cite{talman,talman1} or by expanding the effective
potential in a finite basis set that guarantees the fulfillment of the
correct asymptotic behavior \cite{cone01}. \\

In this work we have generalized this method to deal with multi configuration
wave functions instead a single Slater determinant

\begin{equation}
\Phi_{mc} = \sum_{k} C_k \phi_k
\label{phimc}
\end{equation}

where $\phi_k$ are Slater determinants built with the eigenfunctions of the
effective potential, which has been parameterized in the present
work as follows

\begin{equation}
V(r) = \frac{1}{r} 
\left\{ 
(Z-N+1) + (N-1) \sum_k a_k e^{-b_kr} 
\right\}
\end{equation}

where $Z,N$ are the nuclear charge and the number of electrons,
respectively, and $\{a_k,b_k\}$ are variational parameters with the
constraint $\sum a_k = 1$ to match the exact short range behavior of
the potential. It also reproduces the exact long range behavior. \\

Within this scheme, given some particular values of the parameters
$\{a_k,b_k\}$ the one-body Schr\"odinger equation defined by the
effective potential is solved. The orbitals obtained by solving this
single particle Schr\"odinger equation are used to write down the
Slater determinants appearing in the multi configuration expansion of
the trial wave function of Eq.(\ref{phimc}).  Then the total energy is
calculated as the expectation value of the hamiltonian in that trial
wave function. The linear coefficients $C_k$ of the configuration
interaction expansion of the model wave function are also obtained in
this step. The optimum set of parameters of the effective potential is
fixed by imposing the minimum condition on the total energy.  Finally
it is worth to point out that the one-body problem has been solved
here by expanding the radial part of the single particle wave
functions in terms of Slater functions as in the Roothaan-Hartree-Fock
method.\\

The purpose of the present work is not to obtain the best description
of the excited states of the beryllium atom within the effective
potential framework. It constitutes only a previous step that produces
the model wave functions used along with the correlation Jastrow
factor. Therefore full convergence in the number of free parameters of
the effective potential has not been pursued.  Instead a certain
compromise between the number of free parameters and the accuracy at
the effective potential framework has been found. For example when
using the single configuration $1s^2 2s^2$ to describe the ground
state of the Be atom, the effective potential wave function used here
provides an energy of $-14.57235~au$ to be compared with the optimized
effective potential results $-14.57245~au$ and $-14.57256~au$ reported
in \cite{talman1} and \cite{cone01} respectively and with the
Hartree-Fock value $-14.57302~au$. In the case of a multi
configuration wave function built from the linear combination of the
configurations $1s^2 2s^2$ and $1s^2 2p^2$ that takes into account the
near degeneracy effect and constitutes the complete active space wave
function for this atom, the optimized effective potential energy
obtained by us is $-14.61556~au$ while the exact Multi Configuration
Hartree Fock (MCHF) value with these two configurations is $-14.61685~au$ \cite{davidson}. For the
sake of completeness we mention here that the energy when a Jastrow
correlation factor is used along with the effective potential model
function obtained here is $-14.6647(1)~au$ that coincides with the
value of \cite{galv01} where a MCHF was used as model function, to be
compared with the estimated exact energy $-14.66736~au$
\cite{davidson}. \\

The way in which the different calculations have been carried out is
the following.  For any given a state, for example the $2s3s$--$^1S$,
the set of configurations that are thought to be the most important is
selected, $1s^22s^2$, $1s^22p^2$, $1s^22s3s$ and $1s^22p3p$ in our
example.  With them the model wave function, $\Phi_{mc}$, is fixed
within the effective potential framework. The diagonalization gives
rise to a set of orthogonal and non correlated vectors (four in our
example).  A new correlated basis set is built up by multiplying each
one of the non--correlated vectors by the correlation factor, $F$,
which depends on 17 variational non--linear parameters.  In this new
basis set, which is non--orthogonal, we diagonalize the atomic
hamiltonian, obtaining the eigenvectors along their corresponding
eigenvalues, which are upper bounds to the four first states of
$^1S$--type. Because we are interested in the excited $2s3s$--$^1S$
state, we optimize the non--linear parameters of the correlation
factor by minimizing the second eigenvalue. The minimization has not
been carried out on the variance but on the energy, that has shown to
provide not only more accurate upper bounds \cite{Lin} but also a
better description of other properties \cite{galv01}. The final wave
function, $\Psi_{cmc}$, is written as a correlated basis function
expansion, that in our example is

\begin{equation}
\Psi_{cmc} [2s3s; ^{1}S] = F (d_1 \phi[2s^2; ^{1}S] + d_2
\phi[2p^2;^1S] + d_3 \phi[2s3s;^1S + d_4 \phi[2p3p;^1S]) 
\end{equation}

where the
core $1s^2$ is implicit in each one of the configurations. The
new coefficients $d_k$ are obtained in the 
diagonalization of the Hamiltonian in the correlated basis, and 
the different matrix elements of the hamiltonian in this 
basis set have been calculated by using the Monte Carlo algorithm.\\

{\bf III. RESULTS}\\

The use of a large number of configurations increases greatly the
computer time and also the numerical errors in the
diagonalization. Therefore we have to limit ourselves to consider a
number of configurations as small as possible. Thus, for each one of
the states studied, a given configuration has been selected if the
difference between the energies obtained with and without it is
clearly greater than the statistical error in the calculation, once
the correlation factor has been included. In some cases these
configurations do not coincide with those most important if the
correlation factor, $F$, is not considered.  The best wave functions
obtained in this work recover more than $90 \%$ of the correlation
energy.  The correlation energy has been traditionally defined in
quantum chemistry calculations as the difference between the exact
(non-relativistic, infinite nuclear mass) and the Hartree-Fock
energies. However for some of the states studied in this work the
Hartree--Fock method does not provide an upper bound on the
energy. This problem is handled here by defining the correlation
energy as the difference between the exact non relativistic energy and
that obtained with the best model wave function $\Phi_{mc}$.\\

We have studied the first ten excited states of the beryllium atom, five
singlets and five triplets. The
configurations used to study each one of the states considered are
shown in Table I. In general, configuration mixing is much more
important for singlet than for triplet states. In this table we 
show the excitation energy, in $eV$, obtained for each
one of the states studied. The non--correlated energy obtained
from $\Phi_{mc}$ by 
using these configurations as well as the correlated energy obtained
from $\Psi_{cmc}$ are compared with the estimated exact one
\cite{bash}. In parenthesis we show the statistical error in the Monte
Carlo calculation and, in brackets, the percentage of correlation
energy recovered. For triplet states the difference between the energy
obtained in this work and the exact one oscillates around 0.07 $eV$,
except for the $2s3p$--$^3P$ for which is 0.1 $eV$. For singlet states
this difference goes from 0.1 to 0.14 $eV$.  The greatest difference
is 0.142 $eV$, i.e. 0.005 hartree, and corresponds to the state
$2s3p$--$^1$P.  In general the contribution to a given singlet state
from other configurations than the principal one is much more
important than for a triplet state. As a consequence, the energy
obtained for singlet states within a subspace given is, in general, of
a poorer quality than for triplet states. \\

In Tables II, III and IV we show the results obtained for the moments
of the single--particle and the electron pair intracule and extracule
densities, respectively, starting from the correlated wave function
$\Psi_{cmc}$.  We give the expectation values $\langle t^n \rangle $,
where $t=r, r_{12}$ and $R$, respectively, and $n=-2, -1,1,2$ and
$3$. We have also calculated the value of the single--particle and of
both, the intracule and the extracule, densities at the origin. In
doing so the following relation has been used

\begin{equation}
\chi(0) = - \frac{1}{2 \pi} \sum_{i>j} \int Q \Psi^2(x)
\frac{1}{Q}\frac{1}{ \Psi(x) t^2} \frac{\partial \Psi(x)}{\partial t} dx
\end{equation}

for the single--particle density ($\chi = \rho$, $t=r$) \cite{roth}
and for 
the intracule ($\chi = h$, $t=r_{12})$ and the extracule ($\chi = d$,
$t=R)$ densities \cite{sarsa2}. These expressions allow one to obtain
a local property of the corresponding density in terms of the wave
function evaluated in the whole domain.  In order to calculate all
these expectation values we have used $|\Psi_{cmc}|^2$ as distribution
function, except for $h(0)$ and $\langle r_{12}^{-2} \rangle$ for
which we have used $Q|\Psi_{cmc}|^2$, with $Q= \sum_{i<j} 1/r_{ij}^2$,
which provides more accurate results for those expectation values.\\
 
To study the quality of the results in Tables II and III we compare
the results obtained for the ground state with those than can be
considered as exact obtained also from an explicitly correlated wave
function \cite{komasa}, and with those obtained in \cite{galv01} where
a correlated wave function similar to $\Psi_{cmc}$ but with a MCHF
model wave function was used instead. The results obtained in this
work are of a great quality, improving those reported in
\cite{galv01}. In Tables III--IV we also included the results recently
obtained for the ground state intracule and extracule densities of
beryllium starting from a MCHF wave function \cite{koga3}. Despite
these last results are unaffected of statistical errors and are very
precise in a numerical sense, they are not as accurate as ours at low
interelectronic distances which is the region where electronic
correlations are more important. That shows the validity
of our techniques and our wave function that, although approximated,
accounts for the most relevant physical correlation mechanisms.  In
Tables III--IV we also include the HF results
obtained for the moments of both the intracule and extracule densities
of the $2s2p$--$^1P$ and $2s2p$--$^3P$ states \cite{koga1} and some
values obtained for the intracule density of those same states
starting from a multi configurational wave function \cite{tanaka}. \\

To study the effects of electronic correlations we have considered
those states for which the single configuration Hartree-Fock method can
be applied to obtain upper bounds to the exact non--relativistic
energy of the corresponding state.  These states are $2s2p$--$^3$P,
$2s2p$--$^1$P, $2s3s$--$^3$S , $2p^2$--$^1$D, $2p^2$--$^3$P, and
$2s3d$--$^3$D, i.e. the most bounded state with a $L,S$ given, except
the $2p^2$--$^3$P state because at a Hartree Fock level this state is
orthogonal to the $2s3p$--$^3$P one.  For all these states we have
worked with a monoconfigurational wave function calculated from the
optimized effective potential approach, that it is very similar to the
Hartree-Fock one.  For some other states it is necessary to work with
an interaction configuration wave function that includes partially
correlation effects.\\

The differences found between correlated and Hartree--Fock results are
much more important for singlet than for triplet states.  To analyze
systematically the different correlation mechanism included by the
trial function we have calculated the following difference functions

\begin{eqnarray}
f_{cmc-sc}(t) &=& 4 \pi t^2 [f_{cmc}(t) - f_{sc}(t) ] \\
f_{mc-sc} (t) &=& 4 \pi t^2 [f_{mc}(t)  - f_{sc}(t) ] \\
f_{cmc-mc}(t) &=& 4 \pi t^2 [f_{cmc}(t) - f_{mc}(t) ] 
\end{eqnarray}

where $f$ and $t$ stand for $R$ and $r$, $H$ and $r_{12}$, and $D$ and
$R$, respectively. The $f_{cmc}$ function is the corresponding
density calculated 
from the best correlated wave function obtained in this work,
$\Psi_{cmc}$, $f_{mc}$ correspond to a multiconfiguration wave
function without Jastrow correlation factor, $\Phi_{mc}$, and finally
$f_{sc}$ is the density calculated from the single configuration self
consistent approach wave function.  The $f_{cmc-sc}(t)$ difference
functions account for the full effect of the electronic correlations
on the densities, the $f_{mc-sc}(t)$ give us insight on the effects of
the configuration mixing on the electronic distribution and finally
$f_{cmc-mc}$ give us information on the importance of the Jastrow
factor.\\

In Figure 1 we show the results obtained for $f_{mc-sc}$ and
$f_{cmc-sc}$, for the $2s2p$--$^1P$ and $2s2p$--$^3P$ states which are
representative for those results obtained for singlet and triplet
states. The same is shown in Figure 2 for the $f_{cmc-mc}$ difference
functions.  Several are the effects that can be noticed:\\

i) The big magnitude of the several difference functions for
singlet states as compared with those for triplet ones show that the
monoconfigurational self consistent wave function is not adequate for
describing singlet states and a multiconfigurational wave function is
necessary to include those aspects more relevant in the structure of
these states.\\

ii) For singlet states, the medium and long range behavior of the
three difference functions is well described by the model wave
function, $\Phi_{mc}$. However short range correlations, which are
explicitly included in the correlation factor, $F$, are necessary to
reproduce the Coulomb hole in $h(r_{12})$ as well as the short $R$
behavior of the extracule density.\\

iii) For triplet states the conclusions are not clear: the difference
functions $f_{mc-sc}$ are quite flat, except for the extracule
function, although they follow the correct trend for large values of
the variable and then they include large range correlations. The
functions $f_{cmc-sc}$ show a great amount of structure for all the
three densities. That indicates that for triplet states the
interaction configuration wave function is not so important as in the
case of singlet states.\\

iv) As can be seen in Figure 2, the effect of the correlation factor
on the three difference functions is similar for triplet and singlet
states for short and medium distances, indicating that the correlation
factor is nearly independent of the total spin of the system. The
differences between singlet and triplet become more important at
larger distances.\\

v) The main effects of the electronic correlations take place at low
values of the interparticle distance in the intracule density, the
so called Coulomb hole. The structure of this Coulomb
hole is practically the same for all the states considered in this
work.\\

A comparative study of the states arising from the same
configuration shows that singlet states are more extended in space
than triplet ones. This is due to the outermost electron, as can be
checked by comparing the moments of positive order of singlet and
triplet states in Tables II and III.  This fact is particularly
relevant for the $2s2p$, $2s3d$ and $2s3s$ configurations and
specially important in the $2p^2$ configuration. The $2s3p$
configuration is the case where this effect is less important. We have
verified that the Hartree-Fock approximation, although gives a
qualitatively correct picture of this, provides values that
overestimate this effect.  Because of this more diffuse character of
the electron cloud of the singlet as compared with the triplet one
could think that both electron-electron and electron-nucleus
interactions are stronger in the latter.  However this does not hold
in general, especially for high lying configurations. For example 
the electron--electron repulsion energy, $V_{ee} =
\langle r_{12}^{-1} \rangle $, is higher for singlet than for triplet
states in the configurations $2s3p$ and $2s3d$. Besides, the
electron--nucleus attraction energy, $V_{en} = - Z \langle r^{-1}
\rangle $ is slightly higher for the singlet state than for the
triplet one in the configuration $2s3p$.  \\

The Hund's rule is satisfied for all the configurations studied in
this work except for the $2p^2$ one. In order to shed light over this
fact we have analyzed the balance between the nuclear
attraction $V_{en} = - Z \langle r^{-1} \rangle $ and electron-electron
repulsion $V_{ee} = \langle r_{12}^{-1} \rangle $ energies. For the
$2s2p$ and $2s3s$ configurations the lower energy of the triplet is due
to a tighter nuclear attraction that compensates the electron-electron
repulsion that is also stronger in the triplet. This same
interpretation has been given previously for other systems as
helium-like ions \cite{boyd84} studied by means of correlated wave
functions and some states of the atoms C, N, O, Si, S and others
\cite{koga2} studied by using self consistent type wave functions.
This is the opposite to the traditional interpretation of the Hund's
rule based on the assumption that the repulsion is smaller in the
triplet state. We have found that the only configuration for which
this traditional interpretation holds is the $2s3p$ one, that presents
a stronger repulsion in the singlet that compensates the effect of the
nuclear attraction that is slightly bigger also for the singlet. In the
case of the $2s3d$ configuration both the nuclear attraction and the
electronic repulsion tend to make the triplet more bounded than the
singlet.  Finally, the only configuration that does not verify the
Hund's rule is the $2p^2$ one. This is due to the stronger electronic
repulsion of the triplet that beats the bigger nuclear attraction of
this state as compared to the singlet.\\

A comparative study of the singlet--triplet single--particle,
intracule and extracule densities can be seen in Figure 3 where we
show the difference functions $4 \pi t^2 [f_1(t) -f_3(t)]$ where $t$
stands for $r$, $r_{12}$ and $R$, and $f_1$ ($f_3$) for the singlet
(triplet) single--particle, intracule and extracule densities. In
Figure 3 (top) we show the differences for the single--particle density. At
very short distances  appreciable differences are observed only for the
$2p^2$ and $2s3d$ configurations. For the former the singlet density
is clearly higher than the triplet one. For the $2s3d$ configuration
the triplet shows slightly higher values than the singlet state. For
the other configurations the statistical errors associated to the
Monte Carlo calculation do no allow us to appreciate clearly the
differences between singlet and triplet single--particle density for
short distances.  The configurations with an electron in the M shell
behave differently than the $2s2p$ and $2p^2$ ones at low and medium
distances, where the difference function show a region where it is
positive. That means that for those configurations the singlet states
have a higher density at both low and large electron--nucleus
distances, whereas for the $2s2p$ and $2p^2$ the singlet density is
clearly smaller than the triplet one at low and medium
$r$--distances. It is evident from this figure that the decrease in
the value $\langle r^{-1} \rangle $ for singlet states is a much more
complex electronic redistribution than the one observed for the $2s2p$
\cite{tanaka,koga1}. \\

At the middle of Figure 3  we plot the difference function for the intracule
density. Again the behavior of the $2s2p$ and $2p^2$ difference
functions is very different to the rest of the cases studied. The
first two configurations show a deep minimum at short interelectronic
distances that indicates that, despite the Fermi hole, electrons are
more likely to be at short interelectronic distances in the triplet
state than in the singlet one. However the situation is different in
the other configurations where the electronic radial intracule density
is higher in singlet states for short and large interelectronic
distances. \\

At this point it is important to note the difference in magnitude
between the $2s2p$ and $2p^2$ configurations. The first one shows a
relatively deeper hole for the single--particle than for the intracule
density that allows one to explain the Hund's rule as usual. However
for the $2p^2$ configuration the magnitude of the hole in the
intracule density is much greater than the corresponding one in the
single particle density. This means that the electron--electron
repulsion energy is much smaller in the singlet than in the triplet state,
leading to an inversion in the order of the states. \\

The extracule density give us further insight into these facts. 
For the $2s2p$ configuration, the electrons in the triplet have a
tendency to be at opposite positions with respect to the nucleus to
reduce the increased electron repulsion energy \cite{koga1}. This
effect is much less pronounced in the $2p^2$ configuration as can be
seen in Figure 3 (bottom), while for the rest of the configurations the
opposite holds.\\

The results obtained for the $2s2p$ configuration show the same
behavior as in the Hartree--Fock framework \cite{koga1}, although the
numerical values are different. These two  two--body difference
functions decrease from zero to a minimum value, then increase to a
maximum from where they tends to zero taking positive values.  The
three densities of the $2s3p$ configuration show a minimum at
short distances, not very important in magnitude. This is due to the
different behavior of singlet and triplet densities at their
corresponding second maximum, making this configuration slightly
different from all the rest.\\

Finally, in all cases the difference functions approach to zero at
large distances from positive values, that indicates that singlet
states are more extended in space than triplet ones.\\

To better understand these differences, we show in Figures 4, 5 and 6,
the singlet and triplet single--particle, intracule and extracule
radial densities, respectively for each one of the configurations
studied here.  In all cases the pictures show two or three maxima,
corresponding to the number of shells involved in each
configuration. The third maximum is more difficult to observe in the
single--particle density, as it can be seen in Figure 3, but it is
very important in both the intracule and extracule densities. In all
the cases considered the more important differences are found for the
second and third maxima due to the different coupling of the outermost
electrons.  
For the $2s2p$ and $2p^2$ configurations, the second maximum of the single
particle density is higher for the triplet than for the singlet
state. For the other configurations the second maximum is practically
the same for the singlet and the triplet while the third maximum is
again higher for the triplet than for the singlet.  For both the
intracule and the extracule densities the behavior is similar as in
the single--particle density with some exceptions. For the $2s3s$ and
$2s3d$ configurations in both the intracule and extracule densities
the second maximum of the singlet is higher than its counterpart in
the triplet state. For the $2s3p$ states the second maximum of
the extracule density is higher for the singlet than for the triplet, 
while  the contrary holds for
the intracule density.  The differences
for the single--particle density are less important than for the
intracule or extracule densities.\\

In Table V we report some two body properties in position and momentum
spaces. In particular we give the position expectation value $\langle
\vec{r}_1 \cdot \vec{r}_2 \rangle $ = $\langle \sum_{i>j} \vec{r}_i
\cdot \vec{r}_j \rangle$ and the angular correlation factor, $\tau$,
introduced by Kutzelnigg, Del Re, and Berthier \cite{krb}, defined as

\begin{equation}
    \tau= \frac{2~   \sum\limits_{i>j}
                  \left\langle
\vec{r}_i \cdot \vec{r}_j \right\rangle }
{(N-1) \sum\limits_i   \left\langle  r_i^2  \right\rangle }
\label{taur}
\end{equation}

This quantity is bounded in magnitude by unity, $-1 \leq \tau \leq 1$.
$\tau=1$ ($\tau=-1$) means perfect positive (negative) correlation and
$\tau=0$ is for non--correlated variables. For atomic systems this
coefficient is a measure of the averaged relative angle of the
position vectors of each pair of electrons, and it is related to
quantities that may be obtained from experimental measurements, such
as the diamagnetic susceptibility and the dipole oscillator strength
distribution. Similar definitions are done in momentum space. \\

In position space these results indicate that the states here
considered present negative angular correlation, except for the
singlet from $2s2p$ and $2s3d$. This is consistent with the previous
analysis carried out for the extracule density. In momentum space all
the states show positive angular correlation. We have checked that
these values depend strongly of the configuration mixing chosen in
such a way that a small difference in energy can lead to a difference
of about $5 \%$ for these coefficients.\\

We also report in Table V some correlated momentum expectation values
which are directly obtained in the Monte Carlo calculation. In
particular we show the expectation values $\langle p^2 \rangle$,
$\langle p_{12} ^2 \rangle$, $\langle P^2 \rangle$ and $\langle
\vec{p}_1 \cdot \vec{p}_2 \rangle$ These values are compared with
those obtained from $\Phi_{mc}$, i.e. with those obtained from a
multiconfiguration non explicitly correlated wave function. As it is
known these two body expectation values as well as $\langle p^2
\rangle$ are not linearly independent, but they satisfy some relations
that, for four electron atoms, are written as \cite{galv00}

\begin{equation}
3 \langle p^2 \rangle = \langle p_{12} ^2 \rangle + 2 \langle
\vec{p}_1 \cdot \vec{p}_2 \rangle \\
\end{equation}
\begin{equation}
3 \langle p^2 \rangle = 4 \langle P^2 \rangle - 2 \langle \vec{p}_1
\cdot \vec{p}_2 \rangle \\
\end{equation}

As can be noticed from Table V the effects of electronic correlations
are very important for $\langle p^2 \rangle$ and henceforth for the
kinetic energy. This can be analyzed in terms of the previous
equations. The value of $\langle \vec{p}_1 \cdot \vec{p}_2 \rangle$
from a non correlated wave function is nearly zero, except for the
$2s2p$--$^3$P state and for the $2p^2$ terms, and becomes positive
when correlations are taken into account. Electronic correlations also
increase $\langle P^2 \rangle$ while reduce the value of $\langle
p_{12}^2 \rangle$. This same trend was also found for the ground state
of the isoelectronic series of the beryllium atom
\cite{galv00}. Finally it is worth to stress that the inclusion of the
Jastrow correlation factor reveals to be very important in order to
get a reliable description of these properties.\\

{\bf IV. CONCLUSIONS} \\

The first excited states of the beryllium atom have been studied. We
have considered the singlet and triplet states of the configurations
$2s2p$ ($^3P$ and $^1P$), $2s3s$ ($^3S$ and $^1S$), $2p^2$ ($^1D$ and
$^3P$) $2s3p$ ($^3P$ and $^1P$) and $2s3d$ ($^3D$ and $^1D$).  We have
analyzed the differences between singlet and triplet states and the
effect of electronic correlations on them. This study has been carried
out not only for the energy of the state but also on some other
properties related to the electronic distribution.  Explicitly
correlated wave functions have been used to describe these states.
They include a Jastrow-type correlation factor and a multiconfiguration
model wave function providing accurate values of the correlation
energy. The model wave function has been calculated by a
generalization of the optimized effective potential method to
include configuration interaction type expansions. 
All the calculations of this work have been performed by using
the Variational Monte Carlo Method.\\

The effects of electronic correlations are more important for singlet
than for triplet states. To describe the former it is
absolutely necessary to consider a multi configuration model wave
function for any of the states studied, while for the latter a single
configuration model wave function provides practically the same
results as those given by the multi configuration one.  Both the
correlation factor, $F$, and  the Coulomb hole, show a similar
structure for singlet and triplet states.  The effects of electronic
correlations are more important for the intracule and extracule
densities than for the single-particle density, as one could expect.\\

Singlet states present a more diffuse electronic distribution than
triplet ones. This effect is overestimated in the Hartree-Fock
approximation. This fact is deeply related to the relative binding
energy of the singlet and triplet states because of its influence on
the balance between the nuclear attraction and electron-electron
repulsion energies.  For the $2s2p$ and $2s3s$ configurations we have
found that the bigger magnitude of the nuclear attraction in the
triplet compensates the bigger value of the electronic repulsion of
the triplet giving rise to a lower value of the binding energy,
according to the usual interpretation of Hund's rule.  This is not the
case of the $2s3p$ configuration where the lower energy of the triplet
state is due to the bigger electronic repulsion in the singlet, where
the nuclear attraction is also slightly bigger in magnitude than in
the triplet. For the $2s3d$ configuration both the electron-nucleus
attraction and the electron-electron repulsion contribute in the same
direction to make more bound the triplet state.  The $2p^2$
configuration does not satisfy the Hund's rule because here the bigger
magnitude of the nuclear attraction in the triplet does not compensate
its bigger value of the electronic repulsion.\\

The effect of electronic correlations is also to decrease the
expectation value $\langle \vec{r}_1 \cdot \vec{r}_2 \rangle $,
i.e. to increase the probability of two electrons to be at opposite
positions with respect to the nucleus, except for the singlet $2s2p$
and $2s3d$. In momentum space the situation is the opposite, finding
positive angular correlation. This is related to the increase of the
kinetic energy when correlations are taken into account (virial
theorem) and the pair-center of mass momentum distribution, while the
relative momentum of two electrons becomes reduced.\\

{\bf ACKNOWLEDGMENTS} \\

This work has been partially supported by the Spanish Direcci\'on
General de Investigaci\'on Cient\'{\i}fica y T\'ecnica (DGICYT) under
contract PB98--1318 and by the Junta de Andaluc\'{\i}a.  A.S.
acknowledges the Italian MURST for financial support from the grant
MIUR-2001025498.

\newpage
 
\begin{center}
{\bf
TABLE CAPTIONS} \\ 
\end{center}

{\bf Table I}: Excitation energy (in eV) of the different states of
beryllium obtained from $\Phi_{mc}$ ($E_{mc}$) and $\Psi_{cmc}$
($E_{cmc}$) as compared with those of \cite{bash} ($E_{exact}$). 
In parentheses we give the statistical error in the last digit, and in
bracket is shown the percentage of correlation energy recovered in
each calculation. The configurations used in the model wave function are
also shown with the core electrons $1s^2$ implicit in the notation.\\

{\bf Table II}: Radial moments and value at the origin of the
single--particle density for the different states studied in this
work.  For the ground state the results are compared with those of
\cite{komasa}, which can be considered as exact, and with the VMC of
\cite{galv01}.  For the 2s2p states some HF results are also shown. In
parentheses we give the statistical error in the last digit.\\

{\bf Table III}: Radial moments and value at the origin of the
intracule density.  For the ground state the results are compared with
those of \cite{komasa}, which can be considered as exact, and with
those of \cite{galv01}. For the 2s2p states some HF and CI results are
also shown.  In parentheses we give the statistical error in the last
digit.\\

{\bf Table IV}: Radial moments and value at the origin of the
extracule density. 
For the 2s2p
states some HF results are also shown. 
In parentheses we give the statistical error in the last digit.\\

{\bf Table V}: Several two body position and momentum properties obtained
from $\Psi_{cmc}$ as 
compared with those obtained from  $\Phi_{mc}$  that does not
include the Jastrow factor.
In parentheses we give the statistical error in the last
digit.

\newpage
 
\begin{center}
{\bf FIGURE CAPTIONS} \\ 
\end{center}

{\bf FIGURE 1}: Difference functions between the fully--correlated and
the self consistent results (cmc--sc) and between the
multiconfiguration and self consistent ones (mc-sc) for the radial
single--particle (top), 
intracule (middle) and extracule (bottom) densities. The results for the 2s2p--$^1$P
(2s2p--$^3$P) are shown in the left (right) part of the Figure.\\

{\bf FIGURE 2}: Difference functions $f_{cmc-mc}(t)$ between the fully correlated and
the multiconfiguration radial single--particle (top), intracule
(middle) and extracule (bottom) densities for the $2s2p$ states.\\

{\bf FIGURE 3}: Difference between singlet and triplet radial
single--particle (top), intracule (middle) and extracule (bottom)
densities.\\

{\bf Figure 4}: Comparison between singlet and triplet radial single
particle density for the different excited states calculated from the
best trial wave function obtained in this work.\\

{\bf FIGURE 5}: Comparison between singlet and triplet radial
intracule density for the different excited states calculated from the
best trial wave function obtained in this work.\\

{\bf FIGURE 6}: Comparison between singlet and triplet radial
extracule density for the different excited states calculated from the
best trial wave function obtained in this work.\\

\newpage

\vspace*{-2cm}

\begin{center}

TABLE I\\

\vspace{1cm}

\begin{tabular}{l|l|llllll}
\hline\hline
 & Configurations & $ E_{mc} $ & $E_{cmc}$ & $E_{exact}$ & \\
\hline
 2s$^2$ ($^1$S)    & 2s$^2$;2p$^2$;2s3s &   1.383  &  0.069(1) [95.7]
 & 0 & & \\
\hline
2s2p ($^3$P)  & 2s2p;2s3p;3s2p;2p3d  &   4.147  &  2.795(1) [95.0] &     2.725     & &   \\
\hline
2s2p
($^1$P) & 2s2p;2s3p;2p3d &   6.845  & 5.405(1) [91.8] &   5.277    & &   \\
\hline
2s3s ($^3$S) & 2s3s;2p3p &   7.796   &  6.528(1) [94.7] & 6.457 & & \\
\hline
2s3s ($^1$S) & 2s$^2$;2p$^2$;2s3s;2p3p & 8.245  & 6.875(1) [93.4] & 6.779 & &   \\
\hline
2p$^2$
($^1$D)   & 2p$^2$;2s3d;2s4d;2p3p &   8.460   & 7.157(2) [92.4] & 7.05 & &   \\
\hline
2s3p ($^3$P) & 2s3p;2s2p;3s3p &   8.679  &   7.403(3) [92.7] &   7.303  & & \\
\hline
2p$^2$ ($^3$P) & 2p$^2$;3p2;3d$^2$ &   8.823  & 7.457(2) [96.1]
& 7.401 & &   \\ 
\hline
2s3p ($^1$P) &2s3p;2s2p;3s2p;3s3p;2p3d &   8.926  & 7.604(1)
[90.3] & 7.462 & &   \\
\hline
2s3d ($^3$D) & 2s3d;3s2d;2p3p      &  9.055  &   7.760(1) [95.1]  &  7.694  & &\\
\hline
2s3d ($^1$D) & 2s3d;2p$^2$;2p3p;2s4d  &   9.373   & 8.116(2)
[91.5]  &7.998 & \\
\hline\hline

\end{tabular}

\end{center}

\newpage

\vspace*{-2cm}

\begin{center}

TABLE II \\

\vspace{1cm}

\begin{tabular}{l|llllll}
\hline\hline
 & $\rho(0)$ 
 & $\langle r^{-2}\rangle $ & $\langle r^{-1}\rangle $ 
 & $\langle r\rangle $      & $\langle r^2\rangle $ 
 & $\langle r^3\rangle $       \\ 
\hline 
2s$^2$ ($^1$S) & 35.2(1) & 57.3(2) &
8.427(1) &5.9792(5)& 16.254(3) & 56.55(2)  \\ 
Exact \cite{komasa} & 35.3116 & 57.59808   &
8.42735    &5.97256  & 16.2476   & 56.772    \\ 
\cite{galv01} & 35.3(1) & 57.4(2) &
8.433(2) &5.985(1)& 16.343(9) & 57.43(6)  \\ 
\hline 
2s2p ($^3$P) & 34.7(1) & 56.8(2) & 8.3601(8) & 6.2160(6) & 18.111(4) &
68.81(3)  \\ 
HF  \cite{koga1} &          &      &    & 5.41967  & 17.88637 \\
\hline 
2s2p ($^1$P) & 34.7(1) & 56.6(2) & 8.273(1) & 7.240(1) & 27.83(1) &
152.0(2)  \\
HF  \cite{koga1} &          &           &          & 7.33567   &
37.49063 \\
\hline 
2s3s ($^3$S) & 34.94(9) & 56.6(1) & 8.144(1) & 10.818(2) & 75.03(4)
&705.2(8)  \\ 
\hline 
2s3s ($^1$S) & 35.3(2) & 57.3(2) & 8.1328(9) & 11.804(5) &  95.85(8) &
1067(1)  \\
\hline
2p$^2$ ($^1$D) & 34.8(3) & 56.5(4) & 8.202(1) & 7.983(2) & 36.71(3) &
248.5(4)  \\
\hline
2s3p ($^3$P) & 34.4(1) & 56.1(1) & 8.108(1) & 13.017(4) & 120.66(9) &
1498(2)  \\
\hline
2p$^2$ ($^3$P) & 34.0(2) & 56.1(3) & 8.250(1) & 6.7163(7) & 22.080(6)
&95.81(5)  \\
\hline
2s3p ($^1$P) & 34.9(1) & 56.6(1) & 8.111(1) & 13.686(5) & 142.5(1) &
2020(3) \\
\hline
2s3d ($^3$D) & 35.0(1) & 56.9(2) & 8.093(1) & 12.805(2) & 114.27(2) &
1368(2)\\
\hline
2s3d ($^1$D) & 34.7(2) & 56.7(3) & 8.089(1) & 15.457(7) & 193.7(2) &
3254(5)  \\
\hline\hline

\end{tabular}

\end{center}

\newpage
\begin{center}

TABLE III \\

\vspace{1cm}

\begin{tabular}{l|llllll}
\hline\hline
 & $h(0)$ 
 & $\langle u^{-2}\rangle $  &  $\langle u^{-1}\rangle $   
 &  $\langle u\rangle $      & $\langle u^2\rangle $ 
 & $\langle u^3\rangle $  \\ 
\hline 
2s$^2$ ($^1$S) & 1.6079(8) &
9.542(9) & 4.3741(4)&15.292(1)& 52.97(1) &223.07(8)  \\
Exact \cite{komasa} & 1.60704  &
9.5367  & 4.3747   
&15.272   & 52.854   &222.486         \\
\cite{galv01} & 1.611(1) &
9.55(1) & 4.375(1)&15.305(3)& 53.15(3) &224.7(2)  \\
MCHF \cite{koga3} &          & 9.64232
        & 4.38013 &15.2804  & 52.9295  & 223.015   \\
\hline 
2s2p ($^3$P) & 1.571(1) &
9.37(1) & 4.2974(3) & 16.116(2) & 60.17(2) & 276.7(1) \\ 
CI \cite{tanaka} &           &         & 4.374     & 16.206    &
     &              \\ 
HF \cite{koga1} &           &10.31304 & 4.388274  & 16.04838  & 59.72532 &
           \\ 
\hline 
2s2p ($^1$P) &1.5757(8)& 9.24(1) &
4.1434(3)& 18.261(4) & 82.96(5) & 493.4(6)  \\ 
CI \cite{tanaka} &          &          & 4.206    & 18.894    &
     &          \\ 
HF  \cite{koga1} &          & 10.15692 & 4.122636 & 20.64546  &
111.0888 &           \\ 
\hline 
2s3s ($^3$S) &1.589(3) & 9.05(3) & 3.7094(3)&
29.127(6) &230.3(1) &2264(2)  \\ 
\hline 
2s3s ($^1$S) &1.589(4) & 9.02(2) & 3.6874(7) & 31.96(2) &
292.2(3) & 3365(5)  \\ 
\hline 
2p$^2$ ($^1$D) & 1.548(5) & 9.16(6) & 3.9798(3) & 21.212(6) &
118.4(1) & 875(1)  \\ 
\hline 
2s3p ($^3$P) & 1.592(1) & 9.00(1) & 3.6367(4) & 35.38(1) &
363.3(3) & 4627(6)  \\ 
\hline 
2p$^2$ ($^3$P) & 1.518(1) & 9.16(2) &4.1997(3) & 17.01(2) & 67.56(2) &329.6(2) \\ 
\hline 
2s3p ($^1$P) & 1.588(1) & 9.03(2) & 3.6536(4) & 37.47(1) & 430.6(3) &
6230(9)  \\ 
\hline
2s3d ($^3$D) & 1.586(1) & 8.93(1) & 3.6011(2) & 34.804(7) & 345.7(2) &
4259(4)  \\
\hline
2s3d ($^1$D) & 1.584(1) & 8.99(2) & 3.6058(4) &42.58(2) & 580.9(6) &
9908(16)  \\
\hline\hline

\end{tabular}

\end{center}

\newpage
\begin{center}

TABLE IV \\

\vspace{1cm}

\begin{tabular}{l|lllllll}
\hline\hline
 & $d(0)$ 
 & $\langle R^{-2}\rangle $  &  $\langle R^{-1}\rangle $   
 &  $\langle R\rangle $      & $\langle R^2\rangle $ 
 &  $\langle R^3\rangle $      \\ 
\hline
 2s$^2$ ($^1$S) &16.873(5) &43.17(9) & 9.2717(7)&7.0712(6)& 11.139(2)
&21.077(8)  \\
MCHF \cite{koga3} &          &42.9430  & 9.26995  &7.07675  & 11.1826  
&21.2990    \\
\hline
2s2p ($^3$P) & 16.564(4) & 42.56(6) & 9.1955(6) & 7.3087(7) &
12.124(3) & 24.48(1)  \\
HF \cite{koga1} &           & 42.29016 & 9.09576   & 7.43478   &
12.59922  &                    \\
\hline
2s2p ($^1$P)     &16.461(4)&40.97(7) & 8.4700(6)& 9.201(2) & 21.00(1) &
62.18(6)  \\
HF \cite{koga1} &           & 40.398   & 8.12412  & 10.64142
& 29.16552 &                    \\
\hline 
2s3s ($^3$S) &16.555(4)&40.11(7) & 7.6748(5)& 14.201(3) & 54.98(3) &
266.8(3)  \\
\hline
2s3s ($^1$S) &16.545(5)&40.24(7) & 7.657(2)& 15.653(7) & 70.72(6) &
405.1(6)  \\
\hline
2p$^2$ ($^1$D) & 16.351(5) & 41.1(1) & 8.4431(8) & 9.796(3) & 25.46(2)
& 90.3(1)  \\
\hline
2s3p ($^3$P) & 16.558(5) & 40.14(8) & 7.5191(8) & 17.585(5) & 90.16(6) &
573.6(7)  \\
\hline
2p$^2$ ($^3$P) &16.042(6) & 41.3(1) &8.7389(5) & 8.334(1) &16.230(5)
&38.75(2)  \\
\hline
2s3p ($^1$P) & 16.538(5) & 40.3(1) & 7.5318(8) & 18.556(7) & 106.15(9)
& 767(1)  \\
\hline
2s3d ($^3$D) & 16.475(4) & 39.68(7) & 7.4332(6) & 17.211(4)& 84.96(5)
& 521.9(5)  \\
\hline
2s3d ($^1$D) & 16.526(4) & 40.09(9) & 7.4552(9) & 21.28(1) & 145.3(1)
& 1239(2)  \\
\hline\hline

\end{tabular}

\end{center}

\newpage


TABLE V \\

\vspace{1cm}

\begin{tabular}{l|lllllll}
\hline\hline
 & $\langle \vec{r}_1 \cdot \vec{r}_2 \rangle $
 & $\langle \tau_r \rangle $ & 
   $\langle \vec{p}_1 \cdot \vec{p}_2 \rangle $ 
 & $ \langle \tau_p \rangle $ &
$\langle p^{2 } \rangle $ & $\langle p_{12}^2 \rangle $ & 
$\langle P^2 \rangle $ \\
\hline
 2s$^2$ ($^1$S)\\
$\Psi_{cmc}$  &-2.103(1) &-0.08625(4) & 0.456(1) & 0.01036(3)
 &29.342(8)&87.11(2)& 22.234(6) \\
$\Phi_{mc}$ &-1.714(1) &-0.07004(4) & 0.015(1) & 0.00034(3)
&29.301(7)&87.87(2)& 21.983(5) \\
\hline
2s2p ($^3$P) \\
$\Psi_{cmc}$ & -2.920(1) & -0.10747(4) & 0.260(1) & 0.00594(3) & 29.157(6) & 86.95(2) &
21.998(5) \\   
$\Phi_{mc}$ & -3.051(1) & -0.11350(4) & -0.181(1) & -0.00415(2)& 29.058(8) & 87.54(2) &
21.703(6) \\
\hline
2s2p ($^1$P) \\
$\Psi_{cmc}$ & 0.258(2)&0.00617(4) & 0.459(1) & 0.01057(2) &
28.962(7)& 85.97(2) & 21.951(5) \\ 
$\Phi_{mc}$ & -0.132(2) & -0.00306(4) & 0.031(1) & 0.00071(2) &
28.799(7)& 86.33(2) & 21.615(5) \\ 
\hline 
2s3s ($^3$S) \\
$\Psi_{cmc}$ &-2.578(3) & -0.02291(3) & 0.449(1) & 0.01036(3) & 28.879(8)& 85.74(2) &
21.884(6) \\
$\Phi_{mc}$ & -2.690(3) & -0.02430(3) & 0.013(1) & 0.00029(3) & 28.829(9)& 86.46(3) &
21.628(7) \\ 
\hline
2s3s ($^1$S) \\
$\Psi_{cmc}$ &-2.331(4) & -0.01622(3) & 0.440(1) & 0.01016(3) & 28.858(7)& 85.69(2) &
21.864(5) \\
$\Phi_{mc}$  &  0.048(5) &  0.00032(3)  & 0.006(1) &
0.00015(3) & 28.731(8)& 86.18(2) & 21.551(6) \\
\hline
2p$^2$ ($^1$D)\\
$\Psi_{cmc}$ &-4.140(2) & -0.07518(6) & 0.151(1) & 0.00350(2) & 28.845(9)& 86.23(3) &
21.709(7) \\
$\Phi_{mc}$ & -4.042(2) & -0.07229(6) & -0.2978(9) & -0.00691(2) & 28.738(8)& 86.81(2) &
21.405(6) \\
\hline
2s3p ($^3$P) \\
$\Psi_{cmc}$ & -0.673(4) & -0.00372(2) & 0.425(1) & 0.00985(3) & 28.799(7) & 85.55(2) &
21.812(5) \\
$\Phi_{mc}$ & -0.026(5) & -0.00014(3) & -0.005(1) & -0.00011(3) & 28.641(8) & 85.93(2) &
21.478(6) \\
\hline
2p$^2$ ($^3$P) \\
$\Psi_{cmc}$ &-0.660(1) &-0.01994(3) &0.142(1) & 0.00329(3) & 28.815(7) & 86.16(2)
&21.682(6) \\
$\Phi_{mc}$ &-0.999(1) &-0.03070(4) &-0.282(1) & -0.00658(3) &28.573(8) & 86.28(2)
&21.289(6) \\
\hline
2s3p ($^1$P) \\
$\Psi_{cmc}$ & -1.512(6) & -0.00707(3) & 0.409(1) & 0.00947(3) & 28.803(9) & 85.59(3) &
21.807(7) \\
$\Phi_{mc}$ & -1.212(5) & -0.00584(3) & -0.041(1) & -0.00096(2) & 28.618(8) & 85.94(2) &
21.443(6) \\
\hline
2s3d ($^3$D) \\
$\Psi_{cmc}$ & -1.474(5) & -0.00860(3) & 0.440(1) & 0.01020(3) & 28.774(8) & 85.44(3) &
21.800(6) \\
$\Phi_{mc}$  &   0.011(5) &  0.00006(3) & 0.005(1) & 0.00013(3) & 28.682(8) & 86.04(3) &
21.514(6) \\
\hline
2s3d ($^1$D) \\
$\Psi_{cmc}$ & 0.0953(6) & 0.00033(2) & 0.405(1) & 0.00939(3) & 28.766(7) & 85.49(2) &
21.777(6) \\
$\Phi_{mc}$ & 0.367(6) & 0.00120(2) & -0.050(1) & -0.00115(3) & 28.771(8) & 86.41(2) &
21.554(6) \\
\hline\hline

\end{tabular}


\begin{figure}
\vspace{-2cm}
\hspace{-4cm}
\includegraphics[width=8in,height=10in]{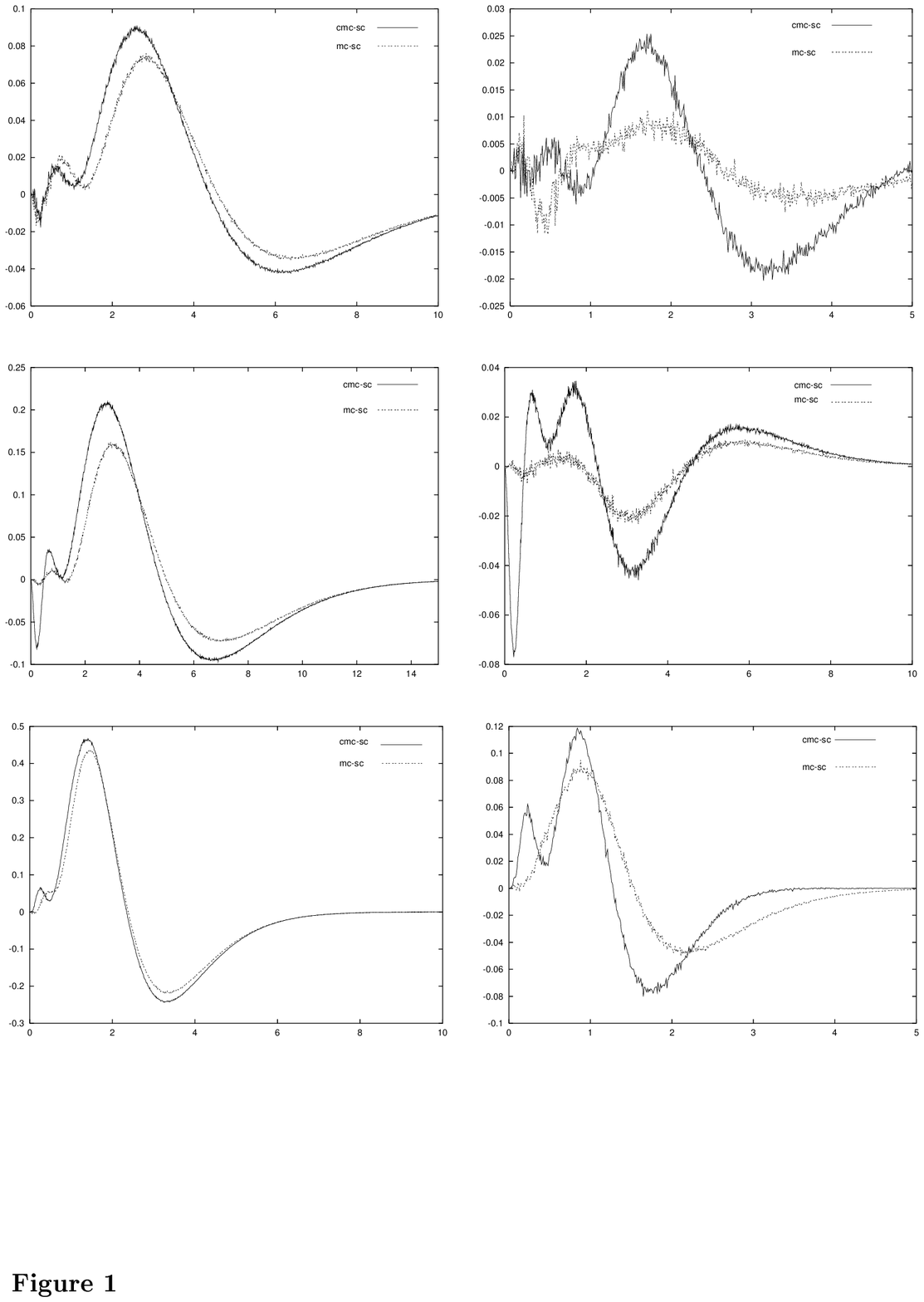}
\end{figure}

\newpage

\begin{figure}
\vspace{-2cm}
\hspace{-4cm}
\includegraphics[width=8in,height=10in]{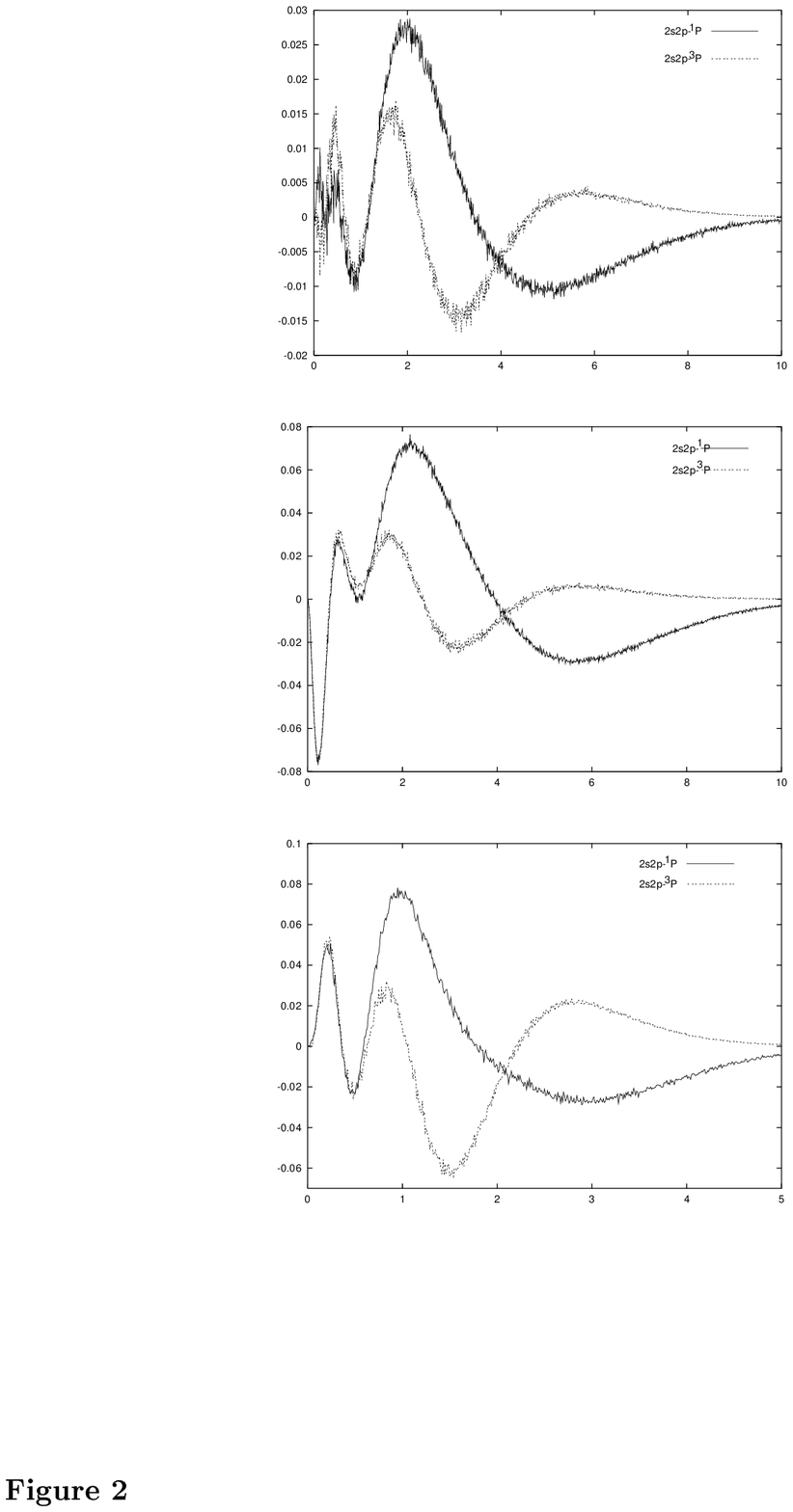}
\end{figure}

\newpage

\begin{figure}
\vspace{-2cm}
\hspace{-4cm}
\includegraphics[width=8in,height=10in]{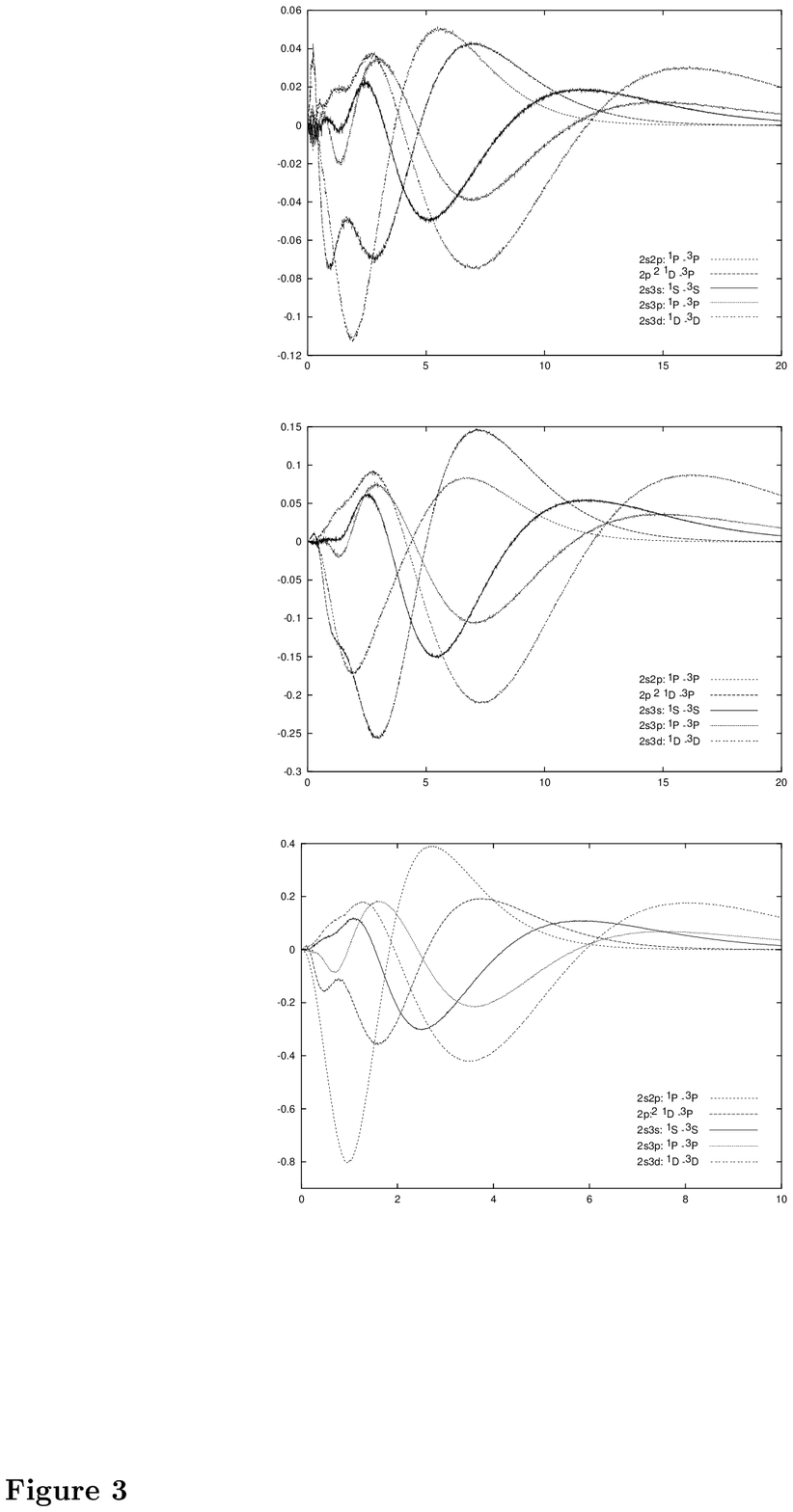}
\end{figure}

\newpage

\begin{figure}
\vspace{-2cm}
\hspace{-4cm}
\includegraphics[width=8in,height=10in]{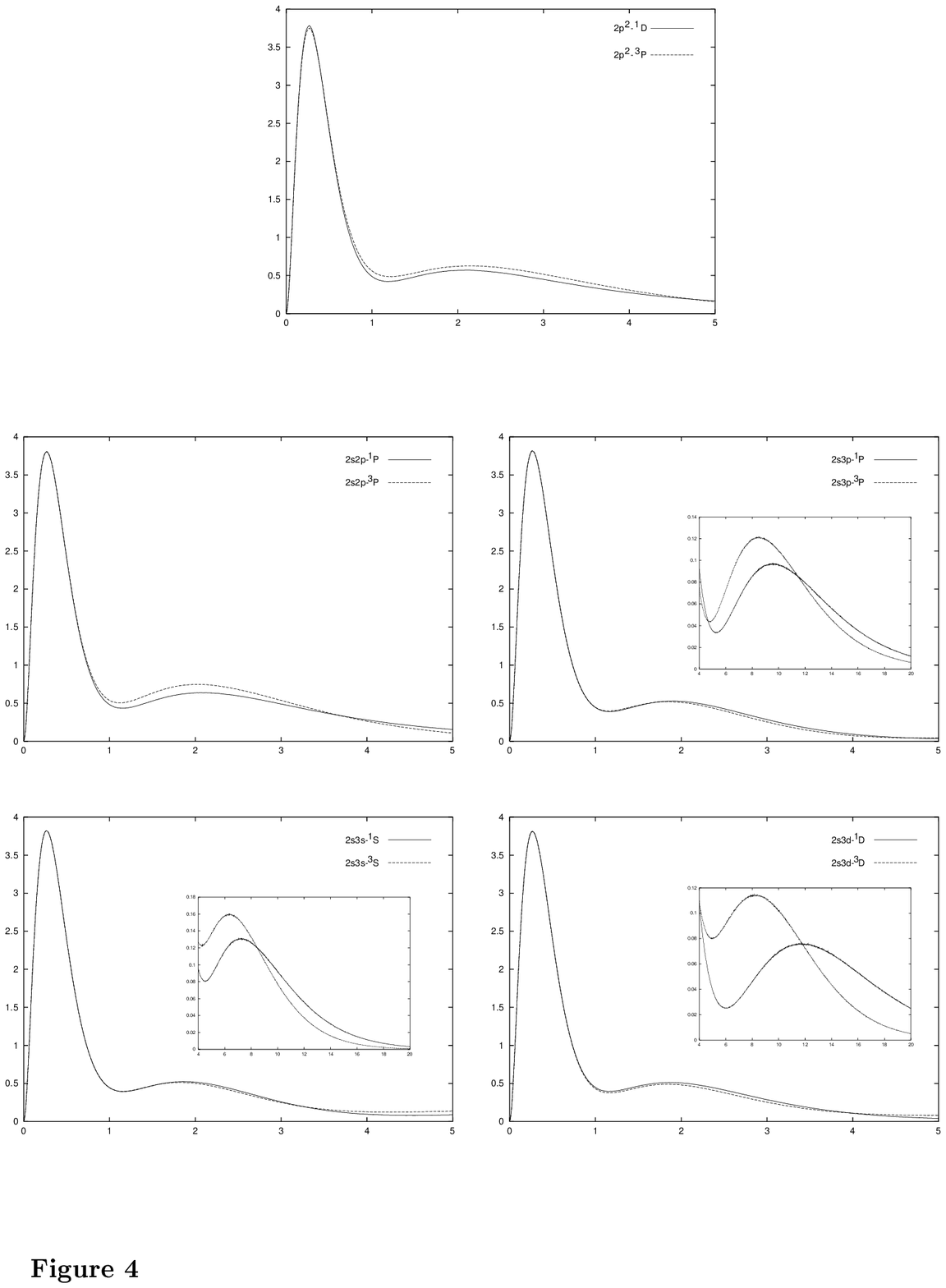}
\end{figure}

\newpage

\begin{figure}
\vspace{-2cm}
\hspace{-4cm}
\includegraphics[width=8in,height=10in]{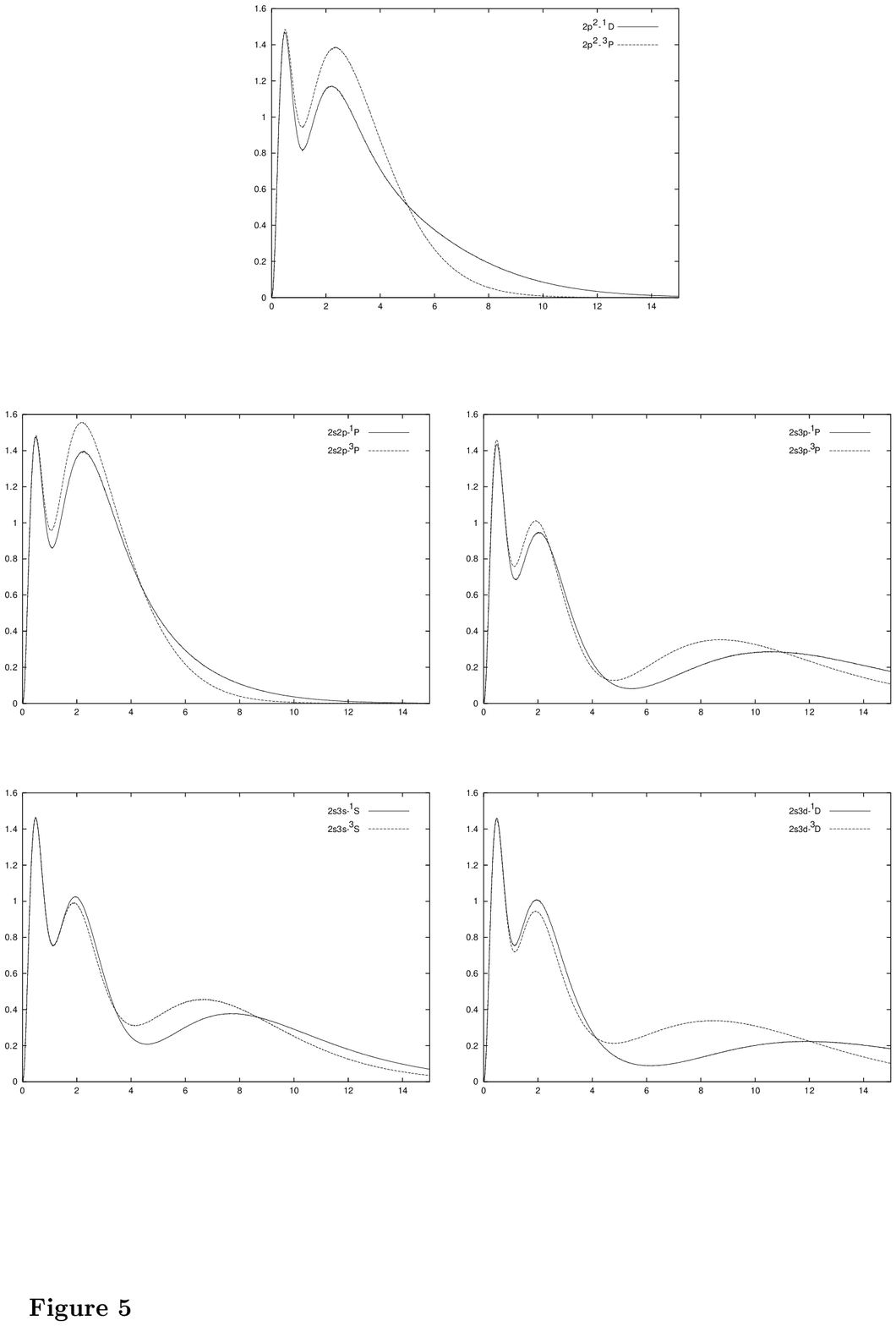}
\end{figure}

\newpage

\begin{figure}
\vspace{-2cm}
\hspace{-4cm}
\includegraphics[width=8in,height=10in]{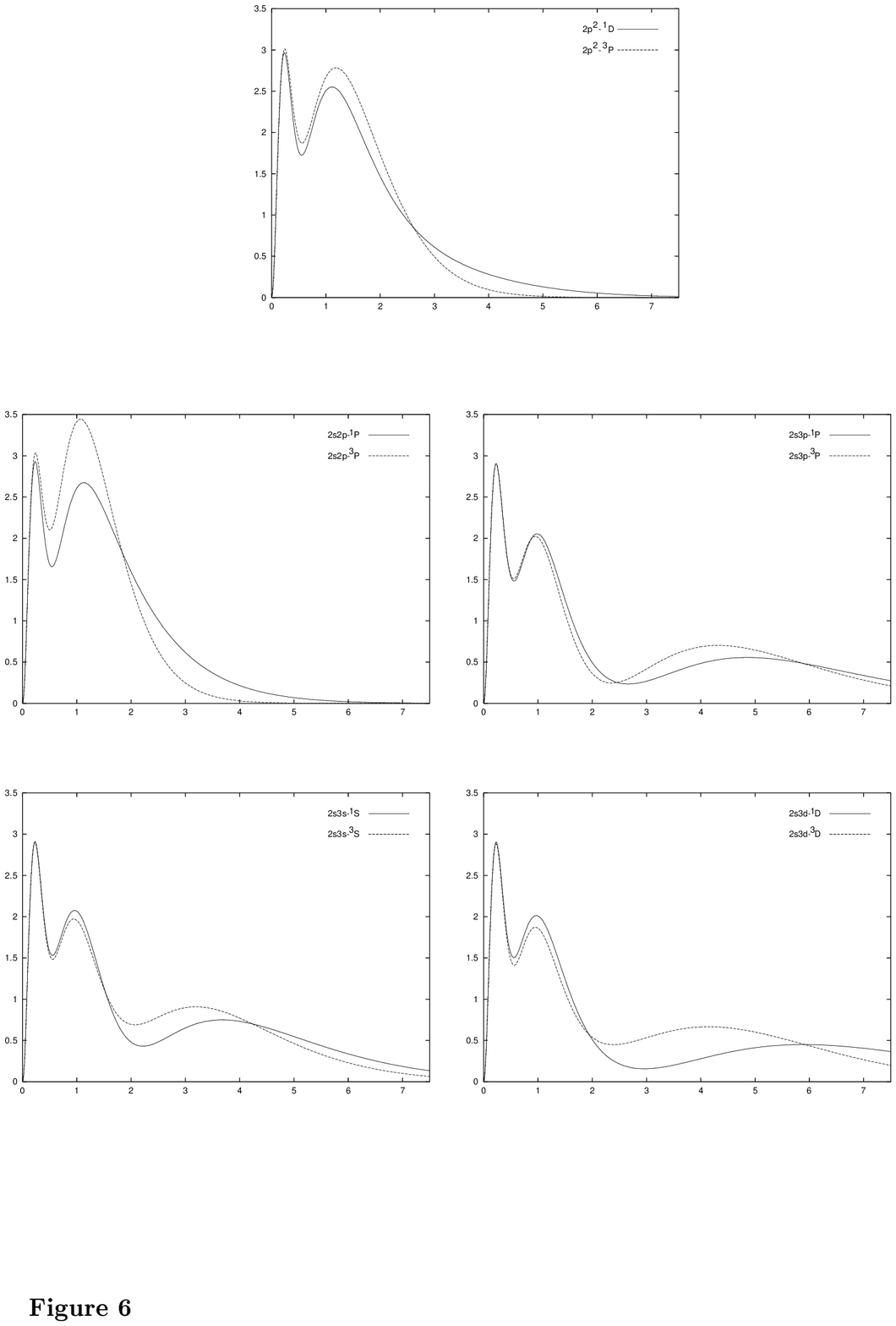}
\end{figure}

\end{document}